\documentclass[twocolumn,showpacs,preprintnumbers,amsmath,amssymb,prb]{revtex4}
\usepackage{graphicx}
\usepackage{dcolumn}
\usepackage{bm}
\usepackage{amsmath,amsfonts,amssymb,amsthm,verbatim,floatflt}
\usepackage{epsfig}
\usepackage[ansinew]{inputenc}
\renewcommand{\i}{\mathrm{i}}

\newcommand{\llangle}{\left\langle}
\newcommand{\rrangle}{\right\rangle}
\newcommand{\vect}[1]{\boldsymbol{#1}}
\begin{document}

\title{The two-impurity Kondo model with spin-orbit interactions} 
\author{David F. Mross\footnote{Current address: Department of Physics, Massachusetts Institute of Technology, Cambridge, MA-02139.} and Henrik Johannesson} 
\affiliation{Department of Physics, University of Gothenburg, SE 412 96 Gothenburg, Sweden}

\begin{abstract}
We study the two-impurity Kondo model (TIKM) in two dimensions with spin-orbit coupled conduction electrons. In the first part of the paper we analyze how spin-orbit interactions of Rashba as well as Dresselhaus type influence the Kondo and RKKY interactions in the TIKM, generalizing results obtained by H. Imamura {\em et al.} (2004) and J. Malecki (2007).  Using our findings we then explore the effect from spin-orbit interactions on the non-Fermi liquid quantum critical transition between the RKKY-singlet and Kondo-screened RKKY-triplet states. We argue that spin-orbit interactions under certain conditions produce a line of critical points exhibiting the same leading scaling behavior as that of the ordinary TIKM. In the second part of the paper we shift focus and turn to the question of how spin-orbit interactions affect the entanglement  between two localized RKKY-coupled spins in the parameter regime where the competition from the direct Kondo interaction can be neglected. Using data for a device with two spinful quantum dots patterned in a gated InAs heterostructure we show that a gate-controlled spin-orbit interaction may drive a maximally entangled state to one with vanishing entanglement, or vice versa (as measured by the concurrence). This has important implications for proposals using RKKY interactions for nonlocal control of qubit entanglement in semiconductor heterostructures. 

\end{abstract}
\maketitle

\section{Introduction}

The study of few-electron quantum dots, or so-called artificial atoms \cite{Kouwenhoven1}, has grown into a major field of research, fueled by the promise of future technological applications, as well as by problems in fundamental physics. Unlike an ordinary atom, the properties of a quantum dot can be manipulated by electrical gates, allowing for well-controlled studies of a whole spectrum of single- and many-particle phenomena, including one- \cite{GoldhaberGordon1} and two-channel \cite{GoldhaberGordon2} Kondo effects, Coulomb blockade \cite{Kouwenhoven2}, and Pauli spin blockade effects \cite{spinblockade}. Recently, nanoscale devices with two spinful quantum dots connected to a conducting region have become the focus of interest, following the breakthrough experiment by Craig {\em et al.} \cite{Craig} (see also Refs. \onlinecite{Sasaki, Heersche}).  When the dots carry spin-1/2 magnetic moments, the problem is that of the {\em two-impurity Kondo model} (TIKM) \cite{Jayaprakash}.

In this model, two localized spins $\boldsymbol{S}_{1,2}$ of magnitude $S=1/2$ are coupled to a sea of conduction electrons via a spin exchange,
\begin{equation}
{H}=H_\text{kin}+J\boldsymbol{S}_1\cdot \boldsymbol{\sigma}_1 +J\boldsymbol{S}_2\cdot \boldsymbol{\sigma}_2.
\label{eqn:isotropickondo}
\end{equation}
Here $\boldsymbol{\sigma}_i =\psi^{\dagger \alpha}(x_i)\boldsymbol{\tau}^\beta_\alpha\psi_\beta(x_i)$ are the electronic spin densities at the sites of the localized spins (with $\boldsymbol{\tau}$ the vector of Pauli matrices), $J$ is the spin exchange coupling, and $H_\text{kin}$ is the kinetic energy of the electrons. To second order in $J$, the conduction electrons mediate a spin exchange interaction between the two localized spins, the Ruderman-Kittel-Kasuya-Yoshida (RKKY) interaction \cite{RKKY}
\begin{equation}
H_\text{RKKY} = K(R)\boldsymbol{S}_1\cdot\boldsymbol{S}_2 ,
\label{RKKY}
\end{equation}
where the strength and the sign of the coupling $K(R)$ depends on the distance $R$ between the spins. 
The competition between the RKKY interaction in (\ref{RKKY}) and the direct Kondo spin exchange in (\ref{eqn:isotropickondo}) is governed by the ratio of $K(R)$ to the Kondo temperature $T_K \sim D \exp(-1/2\rho_F J)$, where $\rho_F$ is the single-electron density of states at the Fermi level. When $\mid\! \!K(R) \! \!\mid \, \gg \, T_K$, the RKKY interaction dominates the direct Kondo exchange and will lock the two impurity spins into a singlet (for $K(R)>0$) or a triplet state (for $K(R)<0$). In the experiment by Craig {\em et al.} \cite{Craig}, the system consisted of two spinful quantum dots in a gated GaAs/AlGaAs heterostructure, coupled to an open conducting region. When tuning the coupling $K(R)$ between the dots, the Kondo signature of the conductance across one of the dots was seen to vanish, or become strongly suppressed, indicating that the RKKY interaction between the dots dominates the Kondo interaction, forming either a singlet state with no Kondo effect or a screened triplet state with a much weaker Kondo effect \cite{SLO,VG}. 

When the strength of the RKKY interaction becomes comparable to the Kondo temperature, $\mid\! \!K(R) \! \!\mid \, \sim T_K$, there is a crossover between the antiferromagnetic RKKY regime (''local singlet'') and the regime where the triplet impurity state is Kondo screened ("Kondo singlet") \cite{Sakai}. If the system possesses particle-hole symmetry, the crossover is expected to sharpen into a second-order phase transition, controlled by a non-Fermi-liquid fixed point \cite{PhysRevB.52.9528}. However, since this symmetry requires a high degree of fine tuning, the possibility to observe the corresponding quantum critical state was for a long time judged as rather unrealistic. However, in recent work by Zar\' and {\em et al.} \cite{zarand:166802} it was shown that the critical state can be stabilized against particle-hole symmetry breaking by using a device where the two dots are connected to two separate leads and RKKY-coupled via a magnetic insulator. 

In a variation of the proposal by Zar\' and {\em et al.}  \cite{zarand:166802}, we here consider a device scheme where the RKKY-coupling is mediated by a Coulomb blockaded auxiliary electron reservoir (see FIG. \ref{fig:fulldevice}). This is the same type of device that was studied in Ref.  \onlinecite{MJ}, with focus on effects from charge fluctuations between leads and dots. This alternative type of setup also allows for the study of effects from {\em spin-orbit interactions} on the critical behavior of the TIKM. This is the main theme of the present paper. As we shall see, the presence of spin-orbit interactions adds a new twist to the problem, with some quite interesting repercussions. Specifically, we shall argue that spin-orbit effects under certain conditions produce a {\em line of critical points}, exhibiting the same scaling behavior as that of the critical TIKM without spin-orbit interactions (up to RG irrelevant scaling corrections).

The role of spin-orbit interactions in the TIKM is also of interest away from criticality, considering proposals for exploiting devices with RKKY-coupled quantum dots in spintronics, with possible applications for future quantum computing. In its most basic implementation, a quantum dot carrying spin-1/2 is here used to represent a qubit,  built from its two spin states $\mid \uparrow \rangle, \mid \downarrow \rangle$ \cite{LD, BLD}.  A key problem is that of connecting the qubits in such a way that their coupling and entanglement become controllable \cite{Cerletti}.  In the original proposal by Loss and DiVincenzo \cite{LD}, the spin qubits are coupled by a Heisenberg exchange, produced by a tunable electrostatic barrier between neighboring dots. For the purpose of computation, however, one must be able to achieve fast and efficient control also of the coupling between 
electron spins on distant quantum dots. A major step towards this goal was taken by Craig {\em et al.} \cite{Craig} in their experimental realization of the TIKM. As shown in the experiment, the coupling $K(R)$ can be turned on and off by the electrical gates that control the energy levels of the dots,
thus making possible a realization of a nonlocal two-qubit logic gate \cite{Ekert}. 
While there are other competing schemes for achieving nonlocal coupling of spin qubits, based on optical 
\cite{Yao} or magnetic \cite{Usaj} control, the RKKY-mediated two-qubit gate has its distinctive advantage in 
being a simple and easily scalable implementation. An estimate by Rikitake and Imamura \cite
{RikitakeImamura}, shows the two-qubit decoherence time (set by the electronic environment that mediates 
the RKKY interaction) to be well within the bounds for efficient gate operations.

In this context, a central question is to understand how robust the RKKY coupling is against competing interactions. In recent work by Cho and McKenzie \cite{cho:012109} it was shown that the entanglement between two RKKY-coupled spin qubits (as measured by the {\em concurrence} \cite{Wootters}) needs a minimum non-zero antiferromagnetic correlation determined by the competition between the RKKY interaction and the Kondo effect. In the second part of this paper we take the Cho-McKenzie analysis \cite{cho:012109} (see also Ref. \onlinecite{Ramsak}) a step further by including the additional effect from spin-orbit interactions. A spin-orbit interaction mixes spin and charge and is known to be an insidious source of decoherence in spin-qubit devices \cite{KhaetskiNazarov}. At the same time, it is possible that one could in fact exploit spin-orbit interactions for coherent control of qubit interactions, as proposed recently by several groups \cite{SOgroups}. We here address the separate issue of how the presence of spin-orbit interactions influence the RKKY coupling and the entanglement between two spin qubits. As we shall see, while the effect may be dramatic, it can be compensated for by properly tuning the electrical gates that define the device, and does not {\em per se} obstruct the operation of an RKKY-mediated two-qubit gate.

The rest of the paper is organized as follows: In Sec. II we review how to derive a modified RKKY interaction in the presence of a spin-orbit interaction of Rashba type \cite{imamura:121303}. The analysis can easily be extended to a spin-orbit interaction of Dresselhaus type, and we carry it out in parallel. The magnitude and sign of the modified RKKY interaction is seen to be strongly sensitive to the strength of the Rashba and Dresselhaus interactions, as encoded in the so called  $\alpha$- and $\beta$-coefficients \cite{Winkler}, with important implications for the design of a spin-qubit device. In Sec. III we then turn to a study of how the Kondo interaction is influenced by spin-orbit interactions of both Rashba and Dresselhaus type, capitalizing on recent work by Malecki \cite{arXiv:0708.2435}.  In Sec. IV we use our results from the previous sections to study the effect of the spin-orbit interaction on the quantum phase transition between the RKKY (''local singlet'') and Kondo screened regimes. In the case when charge fluctuations are present we show that the spin-orbit interaction generates a {\em line of fixed points} (in the language of the renormalization group), and we provide arguments for its interpretation. We also argue $-$ on basis of symmetry arguments $-$ that spin-orbit effects under certain conditions produce a line of critical points (unrelated to the line of fixed points just mentioned), producing the {\em same} leading scaling behavior as that of the critical TIKM without spin-orbit interactions. In Sec. V we change focus and explore how spin-orbit interactions influence the entanglement of two RKKY-coupled spin qubits (in the parameter regime where the direct Kondo exchange can be neglected). As a preamble we derive a general expression for the reduced density matrix and the concurrence for two spin qubits with U(1) symmetry, as appropriate for the present problem where the addition of spin-orbit interactions breaks the SU(2) symmetry of the standard TIKM down to U(1). Sec. VI contains a brief analysis of the behavior of the two-qubit concurrence at criticality, and in Sec. VII, finally, we summarize our results.

\section{RKKY interaction in the presence of spin-orbit interactions} \label{section:RKKY}

\begin{figure}[h]
\centering
\epsfig{file=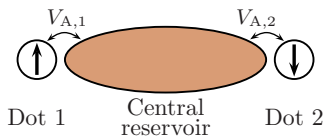, scale=0.9}
\centering
\caption{(color online) The physical system that we study in this section. The different $V$ are tunneling rates. The dots are operated in the Coulomb blockade regime, where charge transfer between the dots and the central reservoir are suppressed.}\label{fig:device}
\end{figure}

In the following we consider a realization of the TIKM in a nanoscale device where two spinful quantum dots are connected via tunnel junctions to a large central electron reservoir. The proposed  device, which may be manufactured using a gated semiconducting heterostructure, is depicted in FIG. \ref{fig:device}. The two dots are coupled to the reservoir via point contacts, allowing the electrons to tunnel between dot $i$ and the reservoir with amplitude $V_{A,i}$.
The dots are operated in the Coulomb blockade regime where transfer of charge between the dots and the reservoir is strongly suppressed but virtual fluctuations give rise to a spin-exchange (Kondo) interaction between the electrons trapped on the dots and the conduction electrons in the reservoir, as described by the Hamiltonian (\ref{eqn:isotropickondo}). With $V_{A,1}=V_{A,2}$, the Kondo couplings to the dots are equal and given by $J \sim V_{A,1}^2/U$, where $U$ is the Coulomb blockade energy of the reservoir. As discussed in the introduction, there exists a parameter regime where the antiferromagnetic RKKY interaction in (\ref{RKKY}), generated by second-order Kondo exchanges, becomes dominant.  As shown in Ref. \onlinecite{cho:012109}, a strong enough antiferromagnetic interaction leads to a finite (in fact, maximal) entanglement between qubits (spins on the dots) as required for a working quantum information device. 
\normalsize

We here focus on effects on the RKKY-interaction from the spin-orbit coupled electrons in the large central reservoir. One may also enquire about the effect from the spin-orbit coupling of the electrons that reside in the quantum dots 1 and 2 (cf. FIG. \ref{fig:device}). While an interesting issue as such, we shall here bypass it by considering the idealized limit of two ultra-small spinful quantum dots, with the trapped electrons modeled as two completely localized spin-1/2 impurities.  In this limit only the spin-orbit coupling of the itinerant electrons in the large reservoir is to be taken into account.

Spin-orbit interactions in semiconductor heterostructures come in two guises, the Dresselhaus \cite{Dresselhaus} and Rashba \cite{Rashba} interactions, originating from the inversion asymmetry of the potential $V(\boldsymbol{r})=V_{cr}(\boldsymbol{r})+V'(\boldsymbol{r})$, where $V_{cr}(\boldsymbol{r})$ is the periodic crystal potential, and $V'(\boldsymbol{r})$ is the potential due to confinement, impurities, external electric fields, etc. The electric field $-\nabla V(\boldsymbol{r})$ produces a Pauli spin-orbit interaction
\begin{equation}
H_{SO} \sim (\boldsymbol{p} \times \nabla V(\boldsymbol{r})) \cdot \boldsymbol{\sigma}
\label{Pauli}
\end{equation}
that can be of sizable magnitude due to the large potential gradients of the atomic cores \cite{Winkler}. Here $\boldsymbol{\sigma} = (\sigma^x, \sigma^y, \sigma^z)$ is the vector of Pauli matrices.
If $V(\boldsymbol{r})$ lacks inversion symmetry, i.e.  $V(\boldsymbol{-r}) \neq V(\boldsymbol{r})$, then the Pauli interaction in (\ref{Pauli}) fails to average to zero in a unit cell, and results in a spin-splitting of the electronic bands.  In semiconductors with bulk-inversion asymmetry (zinc-blende structures, including GaAs) the spin splitting can be encoded in the effective Dresselhaus interaction \cite{Dresselhaus}
\begin{equation}
H_{\text{Dress.}} = {\cal B}\sum_{i\neq j \neq \ell} k_i(k_j^2 - k_{\ell}^2)\sigma^i,
\label{Dresselhaus3D}
\end{equation}
where $k_i, k_j,$ and $k_{\ell}$ are the electronic wave numbers along the principal crystal axes, with $(i,j,\ell)$ cyclic permutations of $(x,y,z)$, and where
${\cal B}$ is a material dependent coupling constant. For a heterostructure grown along $[001]$, with the electrons confined to the $xy$-plane, the Dresselhaus interaction reduces to
\begin{equation}
H_{\beta} = \beta (k_x \sigma^x - k_y \sigma^y)
\label{Dresselhaus2D}
\end{equation}
when taking the average of Eq. (\ref{Dresselhaus3D}) along the z-direction: $\langle k_z \rangle = 0$, and  $\langle k_z^2 \rangle \sim (\pi/d)^2$ with $d$ the characteristic electron wave length in the $z$-direction. Here $\beta$ is the Dresselhaus $\beta$-coefficient, $\beta = - {\cal B}(\pi/d)^2$. 

In a heterostructure, the spin degeneracy can be lifted also because of a structure inversion asymmetry of the confining potential contained in $V'(\boldsymbol{r})$. This potential may obtain contributions also from an externally applied potential as well as from the effective potential from the position-dependent band edges. Assuming that $-\nabla V'(\boldsymbol{r})$ is an electric field along the $z$-direction, one obtains from (\ref{Pauli}) the Rashba interaction \cite{Rashba}
\begin{equation}
H_{\alpha} = \alpha (k_x \sigma^y - k_y \sigma^x),
\label{Rashba}
\end{equation}
where the Rashba coefficient $\alpha$ can be tuned via an external gate
\cite{Nitta, Grundler}. It may be worth pointing out that the spin-orbit interaction comes in two distinct varieties also in other materials. For example, in graphene, a Rashba interaction (\ref{Rashba}) controllable via an external gate electric field coexists with an intrinsic spin-orbit interaction determined by the symmetry properties of the honeycomb lattice, similar to the 2D Dresselhaus interaction (\ref{Dresselhaus2D}) in a semiconductor heterostructure \cite{Paco}. In the following we focus entirely on effects produced by $H_{\beta}$ and $H_{\alpha}$, as defined in Eqs. (\ref{Dresselhaus2D}) and (\ref{Rashba}) respectively.  

Turning to the RKKY interaction, it here acts between two localized spins $S_{1,2}$ in a two-dimensional (2D) electron gas, 
where, in the device in FIG. 1, $S_{i}$ is attached to dot $i \, (i=1,2)$, and the electron gas is represented by the central reservoir. It can be calculated in second order in the Kondo couplings $J_{1,2}$ as \cite{imamura:121303}
\begin{multline} H_\text{RKKY}=
-\frac{J_1 J_2}{\pi}\text{Im}\int_{-\infty}^{\omega_F}
d\omega \ \text{Tr}\, \big[ (\boldsymbol{S}_1\cdot\boldsymbol{\sigma})G({R},\omega+\i 0_+)\\ \times (\boldsymbol{S}_2\cdot\boldsymbol{\sigma})G(-{R},\omega+\i 0_+) \big], \label{eqn:deriverkk}
\end{multline}
with $\omega_F$ the Fermi level.
Here $G({R},\omega)$ is the Green's function of a conduction electron with energy $\omega$, with $2R$ the distance between the localized spins ($\sim$ the width of the central reservoir in FIG. (\ref{fig:device})). The trace is taken over the two spin states of a conduction electron. The expression in Eq. (\ref{eqn:deriverkk}) can easily be adapted to the case when spin-orbit interactions are present by properly modifying the single-electron Greens function $G$. The case of a pure Rashba spin-orbit interaction was treated in Ref. \onlinecite{imamura:121303}. Here we will show that using a similar procedure as in Ref. \onlinecite{imamura:121303}, the form of the RKKY interaction can be calculated in the presence of both Rashba and Dresselhaus interactions of arbitrary relative strength.

For this purpose it is convenient to write the 2D single-electron Hamiltonian with both kinds of spin-orbit interactions (cf. Eqs. (\ref{Dresselhaus2D}) and (\ref{Rashba})) as
\begin{equation}H=\frac{\boldsymbol{k}^2}{2m}+\left[\begin{pmatrix}\beta&-\alpha\\\alpha&-\beta\end{pmatrix} \begin{pmatrix}k_x\\k_y\end{pmatrix}\right]\cdot \vect{\tau}\equiv\frac{\boldsymbol{k}^2}{2m}+\left(\mathbb{A} \boldsymbol{k}\right)\cdot \boldsymbol{\tau},\label{eqn:dressandrash}\end{equation}
where $\alpha$ and $\beta$ are the coupling strengths for Rashba and Dresselhaus spin-orbit interactions, respectively \cite{Rashba,Dresselhaus}. All vectors that appear in Eq. (\ref{eqn:dressandrash}) have two components, i.e. $\boldsymbol{k}=(k_x,k_y)$, $\boldsymbol{\tau}=(\tau^x,\tau^y)$. The scalar product of any vector $\boldsymbol{m}=(m_x,m_y)$ with the vector of Pauli matrices $\boldsymbol{\tau}$ is taken in the usual way, $\boldsymbol{m}\cdot\boldsymbol{\tau}=m_x\tau^x+m_y\tau^y$. The Green's function corresponding to the Hamiltonian in Eq. (\ref{eqn:dressandrash}) is
\begin{align}G(\boldsymbol{k},\omega) & \equiv \left(\omega-H(\vect{k})\right)^{-1}= (\omega-\frac{\vect{k}^2}{2m}-\left(\mathbb{A} \vect{k}\right)\cdot \vect{\tau})^{-1} \nonumber \\ & =G_0(\vect{k},\omega;\mathbb{A})+G_1(\vect{k},\omega;\mathbb{A}) \left(\mathbb{A} \vect{k}\right)\cdot \vect{\tau},
\end{align}
where
\begin{equation}\begin{split}
G_0(\vect{k},\omega;\mathbb{A})&=\frac{\omega -\frac{\vect{k}^2}{2m}}{\left(\omega-\frac{\vect{k}^2}{2m}\right)^2-(\mathbb{A}\vect{k})^2},\\
G_1(\vect{k},\omega;\mathbb{A})&=\frac{1}{\left(\omega-\frac{\vect{k}^2}{2m}\right)^2-(\mathbb{A}\vect{k})^2}.
\end{split}\end{equation}
Note that both $G_0$ and $G_1$ are invariant under the transformation $\vect{k}\rightarrow -\vect{k}$. The Green's function in real space is obtained via a Fourier transform, where
the spin dependence can be pulled in front of the integral by writing
\begin{equation}(\mathbb{A}\vect{k})e^{\i \vect{k}\cdot \vect{R}}=-\i(\mathbb{A}\vect{\nabla})e^{\i \vect{k}\cdot\vect{R}}=\left(\mathbb{A}\hat{R}\right)\left(-\i\frac{d}{d|\vect{R}|}e^{\i \vect{k}\cdot\vect{R}}\right),\end{equation}
with $\hat{R}=\vect{R}/|R|$. We thus obtain 
\begin{equation}G(\vect{R},\omega;\mathbb{A})=G_0(\vect{R},\omega;\mathbb{A})+G_1(\vect{R},\omega;\mathbb{A})\left(\mathbb{A} \hat{R}\right)\cdot \vect{\tau},\end{equation}
where
\begin{equation}G_1(\vect{R},\omega;\mathbb{A})=-\frac{\i}{4\pi^2}\int_{\mathbb{R}^2}d^2k\ \frac{d}{d|\vect{R}|}e^{\i \vect{k}\cdot\vect{R}}G_1(\vect{k},\omega;\mathbb{A}). \label{eq:G1} \end{equation}
Both $G_0(\vect{R})$ and $G_1(\vect{R})$ are invariant under the parity transformation $\vect{R}\rightarrow-\vect{R}$. It is a straightforward task to perform the traces over the Pauli matrices in Eq. (\ref{eqn:deriverkk}) (for details, see Appendix \ref{app:spintraces}), and by specifying a coordinate system where $\hat{R}=\hat{x}$  one obtains from Eqs. (\ref{eqn:deriverkk}) - (\ref{eq:G1}) the RKKY interaction in presence of Rashba and Dresselhaus spin-orbit interactions:
\begin{equation}H^\text{SO}_\text{RKKY}=H_{\text{Heis.}}+H_\text{Rashba}+H_\text{Dress.}+H_{\text{interf.}}, \label{SO} \end{equation}
where
\begin{equation}
 		\begin{array}{clcl}
&H_{\text{Heis.}}&=&F_\text{0}\vect{S}_1\cdot\vect{S}_2\\[1mm]
&H_\text{Rashba}&=&\alpha F_{1}\left(\vect{S}_1\times\vect{S}_2\right)^y+\alpha^2F_2S_1^yS_2^y\\[1mm]
&H_\text{Dress.}&=&\beta F_{1}\left(\vect{S}_1\times\vect{S}_2\right)^x+\beta^2F_2S_1^xS_2^x\\[1mm]
&H_{\text{interf.}}&=&\alpha\beta F_2\left(S_1^xS_2^y+S_1^yS_2^x\right).\label{eq:terms}
\end{array}
\end{equation}
Here $F_i=F_i(\alpha,\beta,R)$ are functions which in the general case are given by rather complicated integrals. For the case where only one type of spin-orbit interaction is present, they have been obtained analytically in the limit of large distances $k_F R\gg 1$ and weak spin-orbit interaction $\alpha,\beta\ll \frac{\hbar^2}{m}k_F$ \cite{imamura:121303}. One finds that \small
\begin{equation}
 		\begin{array}{clcl}
&F_0(0,\alpha,R)&=F_0(\alpha,0,R)&=-\frac{J^2}{2\pi^2R^2}\frac{m^2}{\hbar^2}\sin 2R\sqrt{k_F^2+\frac{m^2\alpha^2}{\hbar^4}}\\[2mm]
&F_1(0,\alpha,R)&=F_1(\alpha,0,R)&=\frac{F_0(0,\alpha,R)}{\alpha}\sin \frac{2mR\alpha}{\hbar^2}\\[2mm]
&F_2(0,\alpha,R)&=F_2(\alpha,0,R)&=\frac{F_0(0,\alpha,R)}{\alpha^2}\left(1-\cos \frac{2mR\alpha}{\hbar^2}\right).
		\end{array}\label{eqn:ffunctions}
\end{equation}\normalsize
Note that we pulled the $\alpha\rightarrow 0$ asymptotics out in Eq. (\ref{eq:terms}), i.e. the functions $F_i$ are finite for $\alpha= 0$.
For $\alpha$ and $\beta$ both nonzero, the integrals must be treated numerically. 

The form of the interaction Hamiltonian in (\ref{SO}) (obtained by choosing a coordinate system where $\hat{R} = \hat{x}$) is convenient for reading off the various special cases of an RKKY interaction with no spin-orbit effects included ($H_\text{Heis.}$), the contribution from a pure Rashba interaction ($H_\text{Rashba}$), the contribution from a pure Dresselhaus interaction ($H_\text{Dress.}$), as well as the contribution to RKKY coming from the interference between the latter two when these are simultaneously present  ($H_\text{interf.}$).  While these expressions are suggestive for physical interpretations there is actually a more useful choice of coordinate system, obtained by choosing the angle between $\hat{R}$ and $\hat{y}$ equal to $-\arctan(\alpha/\beta)$. With this choice $\mathbb{A}\hat{R}=\alpha\left(0,(\alpha^2-\beta^2)\cos\arctan(\alpha/\beta)\right)$, and only $\tau^y$ appears in the Green's function. The interaction then takes the simpler form 
\begin{eqnarray} H_{\text{RKKY}}&=&K_\text{H} \vect{S}_1\cdot\vect{S}_2 +K_\text{Ising}S_1^y S_2^y +K_\text{DM}\left(\vect{S}_1\times\vect{S}_2\right)^y. \nonumber \\ & &  \label{RKKYinter}\end{eqnarray} 
The three terms in (\ref{RKKYinter}) can be identified with the Heisenberg, Ising, and Dzyaloshinsky-Moriya \cite{dzyaloshinski,PhysRevLett.4.228} interactions, respectively. Their coefficients $K_\text{H}, K_\text{Ising}, K_\text{DM}$ depend on the distance between the two spins, the Kondo couplings $J_{1,2}$, and the spin-orbit couplings $\alpha$
and $\beta$. Since $H_\text{RKKY}$ in (\ref{RKKYinter}) manifestly conserves $U(1)$ spin symmetry (spin rotations around the $\hat{y}$-axis), Eq. (\ref{RKKYinter}) is easier to work with than the expression in (\ref{eq:terms}), where it is less obvious how to exploit the U(1) symmetry. It is here notable that the structure of the interaction in Eq. (\ref{RKKYinter}) is the {\em same} as that obtained for a pure Rashba spin-orbit interactions in Ref. \onlinecite{imamura:121303}. In particular, the interference between the two types of spin-orbit interactions ($H_\text{interf.})$ in Eq. (\ref{eq:terms}) does {\em not} produce a new structure in (\ref{RKKYinter}). On the other hand, the parameters $K_\text{H},K_\text{Ising},K_\text{DM}$ do depend differently on $R, \alpha$ and $\beta$ in the general case. It is also worth pointing out that the special case $|\alpha|\!=\!|\beta|$ gives $\mathbb{A}\hat{R}=0$, which means that only the Heisenberg term $\sim\vect{S}_1\cdot\vect{S}_2$ appears in the RKKY interaction, like {\em in the absence of any spin-orbit interaction.} The restoration of the SU(2) symmetry for equal strengths of the Rashba and Dresselhaus couplings has been noted earlier \cite{BAB}, and predicted to give rise to a {\em persistent spin helix} (PSH), a helical spin density wave of infinite lifetime. The emergence of a PSH  was subsequently reported in an experiment on GaAs quantum wells \cite{Koralek}.

In the following we will find it useful to express the RKKY interaction in terms of the parameters $K^y\equiv \frac{1}{2}(K_\text{H}+K_\text{Ising})$ and $e^{\i\theta}K^{\perp}=K_\text{Ising}+\i K_\text{DM}$ as
\begin{equation} {H}_\text{RKKY}=K^y S_1^yS_2^y+\frac{1}{2}e^{\i \theta}K^{\perp}(S_1^z + iS_1^x)(S_2^z - iS_2^x) +h.c.\label{eqn:complexrkky}\end{equation} 
Equivalently, we can express $H_\text{RKKY}$ in terms of a rotated second spin, 
\begin{equation}\vect{S}'_2= e^{ \i\theta S_2^y}\vect{S}_2 e^{-\i\theta S^y_2}\label{eqn:spinrotation},\end{equation}
in which case the RKKY interaction becomes \cite{imamura:121303}
\begin{equation}H_\text{RKKY}=K^{\perp}\vect{S}_1\cdot\vect{S}_2'+(K^y-K^{\perp})S_1^yS_2'^y.\label{eqn:rotatedrkky}\end{equation}
Note that in the special case of Eq. (\ref{eqn:ffunctions}) with only Rashba {\em or} Dresselhaus interaction present, the coefficient of the second term in (\ref{eqn:rotatedrkky}) vanishes and the RKKY interaction in terms of a rotated second spin is purely of the Heisenberg-type. For our purposes, the RKKY interaction in the forms of (\ref{eqn:complexrkky}) and (\ref{eqn:rotatedrkky}) are most suitable and will be used in the following.

\section{Kondo effect with spin-orbit interactions}
\begin{figure}[h]
\centering
\epsfig{file=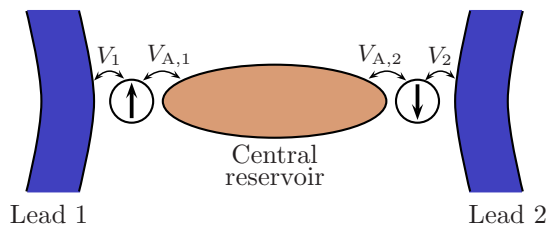}
\centering
\caption{(color online) The physical system that allows for a study of the non-trivial fixed point of the TIKM. The different $V$ are tunneling rates. The dots are operated in the Coulomb blockade regime, where charge transfer between the reservoir and dots as well as between the leads and the dots is strongly suppressed.}\label{fig:fulldevice}
\end{figure}

Having analyzed the effect of spin-orbit interactions on the RKKY coupling between the two localized spins we we must now try to understand how spin-orbit interactions influence the Kondo interaction {\em per se}. The case of Rashba interaction was recently investigated in Ref. \onlinecite{arXiv:0708.2435}, and we here extend the analysis to include also the Dresselhaus interaction.
Since the non-trivial fixed point of the TIKM is unstable against various perturbations (in the language of the renormalization group), we propose a slight modification of the setup that we have discussed so far. The modified setup is depicted in FIG. (\ref{fig:fulldevice}), and has the property that it protects the fixed point \cite{MJ}. \normalsize In the absence of spin-orbit effects, and with the spinful dots operated in the Coulomb blockade regime (so as to suppress charge transfer between dots and leads, as well as between the dots and the reservoir), we can write the effective low-energy Hamiltonian of the device as \begin{equation}
{H}=H_\text{kin}+J\boldsymbol{S}_1\cdot \boldsymbol{\sigma}_1 +J\boldsymbol{S}_2\cdot \boldsymbol{\sigma}_2 + K(R)\boldsymbol{S}_1\cdot \boldsymbol{S}_2.
\label{eqn:isotropickondo2}
\end{equation}
Here $\boldsymbol{\sigma}_i =\psi^{\dagger \alpha}_i(x_i)\boldsymbol{\tau}^\beta_\alpha\psi_{\beta i}(x_i)$ is the electronic spin density in lead $i \, \, (i=1,2,$ with $\boldsymbol{\tau}$ the vector of Pauli matrices). The Kondo coupling $J$ is generated to second order in the tunneling rates, here assumed to be equal, $J \sim V_1^2/U_d$, with $U_d$ the charging energy of the dots. Note that whereas in Eq. (\ref{eqn:isotropickondo}) the spin-densities of the electrons in the central reservoir were coupled to \emph{both} spins, here the spin densities in lead $i$ couple only to the spin $i$. In effect, we have two single-impurity Kondo models coupled via the RKKY interaction $K(R) \boldsymbol {S}_1\cdot \boldsymbol{S}_2$ between the impurity spins. The size of the RKKY coupling $K(R)$ is assumed to be much larger than the Kondo temperature of the reservoir, hence the direct Kondo spin exchanges between the localized electrons on the dots and the electrons in the reservoir have been neglected in Eq. (\ref{eqn:isotropickondo2}). For more details on this, see Refs. \onlinecite{zarand:166802,MJ}.
\normalsize

To find out about the interplay of Kondo, RKKY, and spin-orbit effects, let us begin by writing down the single-impurity Kondo model in two spatial dimensions $(x,y)$ with an added spin-orbit interaction:
\begin{eqnarray}\label{2dkondoso} H\hspace{-.75mm}&=&\hspace{-.75mm}\int d^2k\ \epsilon_k \psi^{\dagger\sigma}_{\vect{k}}\psi_{\vect{k},\sigma}\hspace{-.75mm} \nonumber \\ &+&\hspace{-.75mm}\int d^2k \hspace{-.75mm}\int d^2k'\  \psi^{\dagger\sigma}_{\vect{k}}\vect{\tau}_\sigma^{\sigma'}\psi_{\vect{k}',\sigma'}\cdot\vect{S}\hspace{-.75mm}  \\ &+& \hspace{-.75mm}\int d^2k \hspace{-.75mm}\int d^2k'\ \langle \vect{k},\sigma|H_{s\text{-}o}|\vect{k}',\sigma'\rangle \psi^{\dagger\sigma}_{\vect{k}}\psi_{\vect{k}',\sigma'}, \nonumber \end{eqnarray}
where (cf.  Eqs. (\ref{Dresselhaus2D})) and (\ref{Rashba}))
\begin{equation}H_{s\text{-}o}=(\alpha k_x-\beta k_y)\tau^y+(\beta k_x-\alpha k_y) \tau^x. \end{equation}
Here $|\vect{k},\sigma\rangle=\psi^{\dagger\sigma}_{\vect{k}}|0\rangle$ are momentum eigenstates with $\sigma$ the spin z-component, and $\tau^i$ are the Pauli matrices. As before, $\alpha$ and $\beta$ are the couplings for the Rashba and Dresselhaus spin-orbit interactions, respectively, and we assume a spherically symmetric free-electron dispersion $\epsilon_k$.

The single-impurity Kondo model in the presence of only Rashba spin-orbit interaction ($\beta \!= \!0$) was studied in Ref. \onlinecite{arXiv:0708.2435} where it was found that the qualitative low-energy properties of the model are unchanged by the added interaction. This can be seen by writing the Kondo model with spin-orbit interactions in the form of a two-channel Kondo model, where the different channels of electrons have different couplings to the impurity spin. The spin-orbit interaction in this case is absorbed into the kinetic and Kondo terms of the Hamiltonian and no longer appear explicitly. It is straightforward to apply the same procedure to the Dresselhaus interaction.

As usual it is convenient to expand the electron fields in partial waves around the impurity site, which we here choose to be $\vect{R}=0$. Only the $m=0$ fields participate in the Kondo interaction ($m\in\mathbb{Z}$ being the orbital quantum number), with the kinetic energy being diagonal in $m$. As shown in Appendix \ref{app:sokondo}, the Dresselhaus- and Rashba-type interactions both couple each field to exactly one other field, i.e. a field labeled by $m$ gets connected to a field with either $m+1$ or $m-1$. This simplifies the picture considerably, as can be ascertained by the help of FIG.  \ref{fig:couplings}:
\begin{figure}[h]
\centering
\epsfig{file=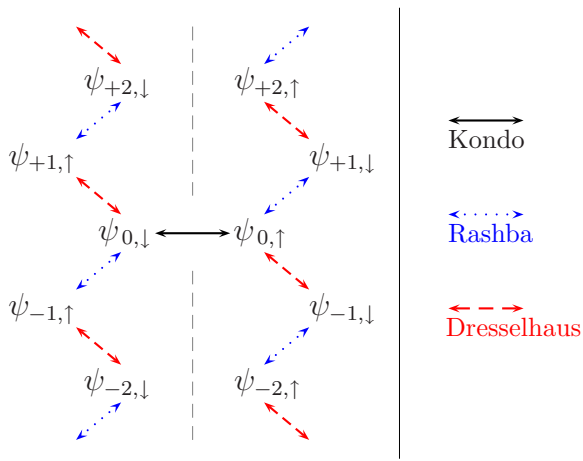, scale=1}
\centering
\caption{(color online) The Kondo interaction as well as both kinds of spin-orbit interactions only couple electron fields with certain quantum numbers, as depicted here. In particular one sees that spin-orbit couplings preserve a ``pseudospin" $\bar{\sigma}$ with the values $\bar{\sigma}=\downarrow$, say, for the fields on the left-hand side and $\bar{\sigma}=\uparrow$ for the ones on the right-hand side. Note that the Kondo interaction only couples fields of different pseudospin.}\label{fig:couplings}
\end{figure}

Since there are no couplings in the kinetic or spin-orbit terms between fields on the left-hand side and right-hand side of the diagram
in FIG. \ref{fig:couplings}, it is possible to choose a basis of simultaneous eigenstates of the kinetic and spin-orbit Hamiltonian, where each state is a linear combination of states from only the left- \emph{or} right-hand side of the diagram, but not from both. For non-zero spin-orbit couplings, each of those fields contains one of the $m=0$ fields and thus couples to the impurity spin.
If both kinds of spin-orbit interactions are present, there is an \emph{infinite} number of fields coupling to the impurity. Unless there is some other mechanism that truncates the angular momentum states that may appear, it is not easy to make a statement about the resulting physics. In fact, an infinite number of fields coupling to the impurity is a similar situation as one would obtain for a long-range Kondo interaction in the absence of spin-orbit couplings.

As long as only one of the spin-orbit interactions is present, the Hamiltonian of Eq. (\ref{2dkondoso}) can be rewritten as
\begin{eqnarray}H&=&\sum_{f,\bar{\sigma}}\int dE E \psi^{\dagger \bar{\sigma}}_{E,f}\psi_{E,f,\bar{\sigma}} 
\nonumber 
\\
&+&\!\frac{1}{2}\int \!dE\int \!dE'
\sum_{\genfrac{}{}{0pt}{1}{f,f'}{\bar{\sigma},\bar{\sigma}'}}
\!J_{f,f'}(E,E')\psi^{\dagger \bar{\sigma}}_{E,f}\vect{\tau}_{\bar{\sigma}}^{\bar{\sigma}'}\psi_{E',f',\bar{\sigma}'} \cdot\vect{S},   \nonumber \\
 \end{eqnarray}
where $f=\pm$ is a flavor index, and $\bar{\sigma}= \downarrow, \uparrow$ is an $SU(2)$ pseudospin-$1/2$ of the electron fields, corresponding to the left- and right-hand sides of FIG. (\ref{fig:couplings}), respectively. The electrons are here expressed in terms of the energy $E$ (for technical details see Appendix \ref{app:sokondo}). This model is known as the (anisotropic) two-channel Kondo model \cite{flavouranisotropy2}. In general, the $J_{f,f'}(E,E')$ interaction is neither diagonal, nor has degenerate eigenvalues. A non-degenerate interaction (whether diagonal or not) is known to drive the two-channel Kondo model towards the single-channel Kondo model (plus one channel of free electrons) under renormalization \cite{flavouranisotropy2}. The physical explanation for this is that if one channel of electrons couples more strongly to the impurity than the other, screening is fully achieved by that channel in the low-temperature limit. In this case, the other channel decouples and behaves like one of free electrons. Therefore, the effective low-energy model describing single-channel Kondo exchange in the presence of spin-orbit interaction of either Dresselhaus or Rashba type, is the usual single-channel Kondo model (without any further interaction), plus a channel of free electrons which decouples and may thus be dropped.

The same argument can be applied to the TIKM. Physically, this is easiest to understand in terms of the nanoscale device introduced in the beginning of the section. Here, the TIKM is understood as two single-impurity models, coupled only via the RKKY interaction. In this picture it is clear that spin-orbit effects should not change the critical behavior. One can, for instance, start with two decoupled single-impurity models, each of which flows to the single-channel fixed point. Then one adds the RKKY coupling, which does not affect how the leads couple to the impurities \cite{FootnoteOnOrder}. The TIKM in the presence of either type of spin-orbit interactions, which is equivalent to an anisotropic two-channel TIKM, can therefore still be described by the Hamiltonian in Eq. (\ref{eqn:isotropickondo2}) as far as the low-energy behavior is concerned.

\section{Criticality in the presence of spin-orbit interactions}\label{sec:sofix}
Before discussing the effects of spin-orbit interactions at criticality of the TIKM, let us briefly review the most important results for the case when no spin-orbit interactions are present.
In this case, an unstable fixed point has been found to exist for a particular value of the parameters, i.e. $K \approx 2.2 T_K$ \cite{PhysRevB.40.324}. At this fixed point the impurity spins form a degenerate doublet between the antiferromagnetic singlet state $|\hspace{-1mm}+\hspace{-.75mm}-\hspace{-.25mm}\rangle-|\hspace{-1mm}-\hspace{-.75mm}+\hspace{-.25mm}\rangle$ and one of the triplet states $|\hspace{-.75mm}+\hspace{-.5mm}+\hspace{-.25mm}\rangle+|\hspace{-.75mm}-\hspace{-.5mm}-\hspace{-.25mm}\rangle$ \cite{gan}, where we label all states in terms of eigenstates of the $y$-component of the spin operators, i.e. $S_i^y | \pm \rangle _i = \pm \frac{1}{2}| \pm \rangle _i$, $i=1,2$. In general, the fixed point is unstable against breaking of particle-hole or parity symmetry \cite{PhysRevB.52.9528}. If charge transfer between the dots is suppressed, the breaking of those symmetries become irrelevant \cite{zarand:166802}. Even {\em with} charge transfer present, breaking the spin $SU(2)$ down to $U(1)$ with an Ising-type interaction $\sim S_1^yS_2^y$ produces only irrelevant operators, while a Dzyaloshinsky-Moriya type interaction allows a marginal operator \cite{footnote3}. The Dzyaloshinsky-Moriya interaction is not invariant under discrete rotations of $\pi$ around the $x$- or $y$-axis, thus this operator may appear in that case.

With this, the stage is set to treat spin-orbit interactions around the fixed point. Since spin-orbit effects give rise to a two-channel model, which in turn renormalizes to a single-channel model, and the RKKY interaction changes as derived in Sec. \ref{section:RKKY}, the effective Hamiltonian can be written as:
\begin{align}H=&H_\text{kin}+J\vect{S}_1\cdot\vect{\sigma}_1+J\vect{S}_2\cdot\vect{\sigma}_2\nonumber\\
&+\underbrace{K^y S_1^yS_2^y+\frac{1}{2}e^{\i \theta}K^{\perp}(S_1^z + iS_1^x)(S_2^z - iS_2^x) +h.c.}_{H_\text{RKKY}}\nonumber \\ \label{eqn:tangle:isokondorkky}\end{align}
 As pointed out above, the Kondo screening behavior is not affected qualitatively; under renormalization the system flows to the same fixed point as without spin-orbit interactions. What may change is the value of the Kondo temperature $T_K$ \cite{arXiv:0708.2435}. Since small deviations around the critical value of $K\equiv K^\perp=K^y\approx 2.2 T_K$ are relevant, changing the Kondo temperature while keeping everything else fixed may drive the system away from the critical point \cite{NotCritical}. To keep the system at the fixed point it may therefore be necessary to fine tune the interactions, which is in principle possible in the proposed nanoscale device. From here on, we assume any changes in $T_K$ to be compensated by modifying $K$ accordingly.
The only way how spin-orbit interactions can then influence the critical behavior is by the symmetry-breaking $SU(2)\rightarrow U(1)$, by which new operators may appear. In terms of the RKKY interaction in Eq. (\ref{eqn:complexrkky}), both ($K^y\neq K^\perp,\theta=0$) and $(\theta\neq 0,K^\perp=K^y)$ break $SU(2)$ down to $U(1)$. We will refer to the former as a longitudinal anisotropy, while we call the latter a transversal anisotropy. As we shall see, the effect of this symmetry breaking depends on the presence of charge transfer between the two channels of conduction electrons \cite{ParticleHole}. There are various cases to consider. We begin with the simplest case of purely transversal perturbations: \\

\paragraph{Transversal anisotropies without charge transfer.}
If $K^\perp=K^y$ the RKKY interaction of Eq. (\ref{eqn:rotatedrkky}) takes the simple form 
\begin{equation}H_\text{RKKY}=K\vect{S}_1\cdot\vect{S}_2'.\end{equation}
We rotate the spins of the conduction electrons coupled to the second impurity spin to match the rotation of the impurity spin (Eq. (\ref{eqn:spinrotation}))
\begin{equation}\psi_2'=e^{-\i\theta \frac{\tau^y}{2}}\psi_2.\end{equation}
The kinetic energy is invariant under these transformations, as are $\vect{S}_1$ and $\psi_1$. It follows that the Hamiltonian takes the form 
\begin{equation}H=H_\text{kin}+K(R) \vect{S}_1\cdot\vect{S}_2'+J\vect{S}_1\cdot\vect{\sigma}_1+J\vect{S}_2'\cdot\vect{\sigma}'_2.\label{eqn:konsoisoagain}\end{equation}
Without the primes this is precisely the same Hamiltonian as for the original (isotropic) TIKM (cf. Eq. (\ref{eqn:isotropickondo2})).
This result does \emph{not} mean that changing $\theta$ is an irrelevant perturbation under which the system flows back to the isotropic fixed point. Instead, the fixed points for all values of $\theta$ should be identified since they arise from the same Hamiltonian. It should be clear that a charge transfer term $\sim\psi^\dagger_1\psi_2$ is not invariant under this transformation, indicating that something different may happen in that case. We will return to this issue later, after considering the case of purely longitudinal anisotropies. Before proceeding, it may be worth noticing that only the singlet state is affected by the transformation above,  turning it into $|\hspace{-1mm}+\hspace{-.75mm}-\hspace{-.25mm}\rangle-\frac{K}{|K|}|\hspace{-1mm}-\hspace{-.75mm}+\hspace{-.25mm}\rangle$. It is  easy to verify that this gives the expected spin-spin expectation value of $\llangle \vect{S}_1\cdot\vect{S}'_2\rrangle=-\frac{1}{4}$. In the original basis this means that the unrotated spin-spin expectation value can take any value between $-\frac{1}{4}$ and $+\frac{1}{4}$. \\

\paragraph{Longitudinal anisotropies with or without charge transfer.}
If $K^y\neq K^\perp$ and $\theta=0$, the Hamiltonian cannot simply be reduced to the isotropic one, even in the absence of charge transfer between the two leads. However, this perturbation is known to be irrelevant from the boundary conformal field theory (BCFT) solution in Ref.  \onlinecite{PhysRevB.52.9528}, even in the presence of charge transfer. In fact, breaking $SU(2)$ to $U(1)$ in this way does not produce a new \emph{leading} irrelevant operator either, which means that for longitudinal anisotropies the system flows back to the isotropic fixed point under renormalization and the scaling behavior of thermodynamic quantities is unaffected.
It should be noted that despite of this, changing $K^y$ while keeping $K^\perp$ constant is a relevant perturbation. This is due to the fact that the shift $(K^\perp,K^y)\rightarrow (K^\perp+\delta,K^y+\delta)$, where $\delta$ is some small number, drives the system away from the fixed point $K^\perp=K^y\approx 2.2 T_K$, and is hence a relevant perturbation \cite{PhysRevB.40.324}. In the linearized renormalization-group 
flow around the fixed point, this means that the irrelevant direction must be perpendicular to this. Only $K^\perp,K^y\rightarrow K^\perp+\delta,K^y-\delta$ can be irrelevant, i.e. $K^\perp$ and $K^y$ need to be changed judiciously in order for the system to remain at the fixed point.\\
\paragraph{Both longitudinal and transversal anisotropies, no charge transfer.}
As long as there is no charge transfer between the leads, the two previous results can be easily combined to determine what happens when both longitudinal and transversal anisotropies are present: By the unitary transformation, the Hamiltonian of such a system reduces to a Hamiltonian where there are only longitudinal anisotropies present. These are irrelevant as noted in the previous case. \\
 
\begin{figure}
\epsfig{file=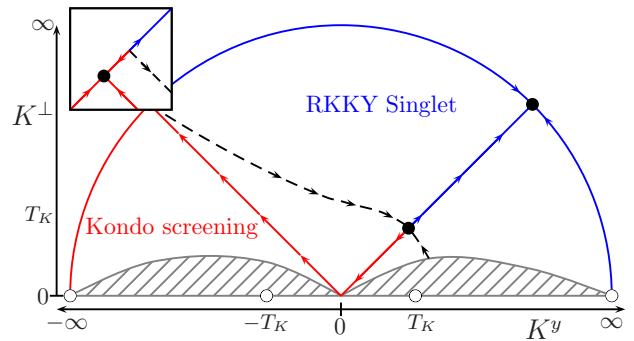, scale=0.9}
\caption{(color online) Qualitative RG flow of the anisotropic model. The solid dots and the solid line are known  results for the isotropic model \cite{PhysRevB.40.324}. The gray area marks the system close to a different  model of quantum dots coupled via an Ising interaction where different behavior is expected \cite{isingcoupled}. The dashed line which we argue for in the text separates the RKKY singlet from the Kondo screened regime.  As an artifact of the scale at $|K|\rightarrow \infty$ it appears curved to coincide with the screened fixed point ($K^\perp=-K^y$). Note that at both the singlet and the screened fixed point, the direction along the semicircle is irrelevant. We thus expect there to remain a finite separation (its scale being set by $T_K$)  between the relevant flow towards the screened fixed point and the dashed flow towards the critical point, as  shown in the enlarged inset.
The curvature of the dashed line is not meant to suggest any deeper knowledge about its properties; however, close to the isotropic (unstable) fixed point it follows the direction of irrelevant longitudinal anisotropies.}\label{fig:flow}
\end{figure}

\paragraph{Both longitudinal and transversal anisotropies as well as charge transfer.}
As pointed out before, the presence of charge transfer disallows the unitary transformation that we used in the previous cases. Exploiting the BCFT solution in Ref. \onlinecite{PhysRevB.52.9528}, we find that an exactly marginal operator ($h_1h_2\phi^y$ in the BCFT language) is allowed when both transversal anisotropies and charge transfer are present. The isospin component $h_1 h_2$ of this operator is only allowed when the charges of the two channels are \emph{not} separately conserved. The spin component $\phi^y$ is allowed if the spin-symmetry is broken down to $U(1)$ \emph{and} the symmetry under discrete rotations by an angle $\pi$ around the $x$-axis and the $z$-axis is broken as well. This is only the case if $\theta\neq 0$, i.e. in the presence of transversal anisotropies. 
A charge transfer term between the two leads $\sim \psi_1^\dagger \psi_2+\psi_2^\dagger \psi_1$, along with the Kondo coupling between the impurity spins and the leads gives rise to an \emph{additional} RKKY-interaction between the two impurity spins, on top of the interaction mediated by the central reservoir. The marginal operator does not drive the system away from criticality, i.e. the RKKY couplings $K^y$ and $K^\perp$ are unchanged (changes in their values are either relevant or irrelevant, see FIG \ref{fig:flow}). Instead, we expect the phase $\theta$ to be replaced by an effective phase $\theta_\text{eff}$, made up of contributions by $\theta$ and by the coupling to the charge transfer terms, which appears as $J_-$ in \onlinecite{PhysRevB.52.9528}. This is consistent with the fact, that the marginal operator is allowed only for $\theta\neq 0$. The models for different values of $\theta$ can therefore no longer be identified as before. This suggests that there is now instead a line of fixed points parametrized by $\theta_\text{eff}$, and connected via the marginal operator.\\ 

\paragraph{Further away from the critical point.}
It should be kept in mind that the notion of an operator being irrelevant only carries meaning in the neighborhood of a fixed point. For larger differences between $K^y$ and $K^\perp$ the situation may be very different. In fact, for $K^\perp=0$ an entirely different quantum phase transition is expected to occur at a particular value of $K^y$ \cite{isingcoupled}. On the other hand, for $K^\perp \gg 0$ and general $K^y<0$, the situation is very much the same as for $K^\perp=K^y$: For sufficiently large values of $K^\perp$, the impurities form an RKKY singlet, while for $K^\perp<-K^y$ the system enters a phase where the two impurity spins are locked in an $S=1$ state and get screened by two effective channels of conduction electrons \cite{flavouranisotropy2}. This suggests that for any given $K^y<0$, there is a value $K^\perp_\text{crit}$ where a phase transition similar to the $K^\perp=K^y$ case occurs. In fact, in the presence of particle-hole symmetry a phase transition \emph{must} occur (see the discussion in Ref. \onlinecite{PhysRevB.52.9528}, p. 9530). Furthermore, the same line of arguments as used in Ref. \onlinecite{gan} to solve the TIKM can be applied in this case (note however, that there it was supplemented by nRG and BCFT results). Since there is no other energy scale in the system, we expect that there is no additional phase transition away from $K^\perp=0$. As illustrated in FIG. \ref{fig:flow}, this suggests that the irrelevant flow to the isotropic fixed point originates from an unstable fixed point at $|K|\rightarrow\infty$ which separates the singlet RKKY and Kondo screened phases.

\section{Entanglement of two RKKY-coupled spins in the presence of spin-orbit interactions}

Having explored how spin-orbit interactions influence the critical behavior of the TIKM, we now shift focus and turn to the question how these same spin-orbit interactions affect the entanglement between the two localized spins deep in the RKKY regime (where the competition from the direct Kondo interaction can be neglected). As discussed in the introduction, this is an important problem considering proposals \cite{LD, BLD} for using RKKY-coupled quantum dots as entangled qubits for quantum computing. 

As a measure of entanglement we shall employ the standard {\em concurrence of formation} \cite{Wootters},  from now on simply referred to as {\em concurrence}. To obtain an expression for the concurrence in the present case it will be helpful to have access to the most general form of a reduced density matrix for a two-spin system compatible with conservation of the $\hat{y}$-component of the total spin. We can write it in terms of six real parameters $a_1,\ldots, a_6$ as \normalsize
\begin{equation}
\begin{array}{rrll}
\rho&=&a_1|\hspace{-1mm}+\hspace{-.75mm}+\hspace{-.25mm}\rangle\langle\hspace{-.25mm}+\hspace{-.75mm}+\hspace{-1mm}|+a_2|\hspace{-1mm}-\hspace{-.75mm}-\hspace{-.25mm}\rangle\langle\hspace{-.25mm}-\hspace{-.75mm}-\hspace{-1mm}|\\[2mm]
&+&a_3|\hspace{-1mm}+\hspace{-.75mm}-\hspace{-.25mm}\rangle\langle\hspace{-.25mm}+\hspace{-.75mm}-\hspace{-1mm}|+a_4|\hspace{-1mm}-\hspace{-.75mm}+\hspace{-.25mm}\rangle\langle\hspace{-.25mm}-\hspace{-.75mm}+\hspace{-1mm}|\\[2mm]
&+&\left(a_5+ \i a_6\right)|\hspace{-1mm}+\hspace{-.75mm}-\hspace{-.25mm}\rangle\langle\hspace{-.25mm}-\hspace{-.75mm}+\hspace{-1mm}| \\[2mm]
&+&\left(a_5- \i a_6\right)|\hspace{-1mm}-\hspace{-.75mm}+\hspace{-.25mm}\rangle\langle\hspace{-.25mm}+\hspace{-.75mm}-\hspace{-1mm}|.
\end{array}\label{eqn:matrix}
\end{equation}
Note that only five of these parameters are independent, due to normalization. We can therefore express the density matrix in terms of the five expectation values
\begin{equation}
\begin{split}
3I&\equiv\langle S^z_1S^z_2\rangle-\langle S^y_1S^y_2\rangle\\
D&\equiv\langle S^z_1S^x_2\rangle-\langle S^x_1S^z_2\rangle\\
3H&\equiv-\langle \vect{S}_1\cdot\vect{S}_2\rangle\\
X^\pm&\equiv\langle S^y_1\rangle\pm\langle S^y_2\rangle.\end{split}\label{eqn:parameters}
\end{equation}
The way the parameters in Eq.  (\ref{eqn:parameters}) are defined allows for an easy reduction to various limiting cases. For example, $H\neq 0, I=D=X^\pm=0$ corresponds to the $SU(2)$ case with a pure Heisenberg interaction between the spins.
Actually, in the system we consider, the parameters $X^\pm$ are always zero by symmetry, i.e. there is no spontaneous magnetization, neither uniform nor staggered. Whereas in the derivation of our formula for the concurrence  (App. \ref{sec:concurrence}),  we keep $X^\pm$ in order to obtain a result that is valid also for more general situations, e.g. in the presence of a magnetic field, we shall drop $X^\pm$ in the following so as to lighten the notation.

We shall also find it convenient to work in the ``magic basis'',  given by
\begin{equation}\begin{split}
|e_1\rangle&=|\hspace{-1mm}+\hspace{-.75mm}+\hspace{-.25mm}\rangle+|\hspace{-1mm}-\hspace{-.75mm}-\hspace{-.25mm}\rangle\\
|e_2\rangle&=\i|\hspace{-1mm}+\hspace{-.75mm}+\hspace{-.25mm}\rangle-\i|\hspace{-1mm}-\hspace{-.75mm}-\hspace{-.25mm}\rangle\\
|e_3\rangle&=\i |\hspace{-1mm}+\hspace{-.75mm}-\hspace{-.25mm}\rangle+\i|\hspace{-1mm}-\hspace{-.75mm}+\hspace{-.25mm}\rangle\\
|e_4\rangle&=|\hspace{-1mm}+\hspace{-.75mm}-\hspace{-.25mm}\rangle-|\hspace{-1mm}-\hspace{-.75mm}+\hspace{-.25mm}\rangle.\label{eqn:magicbasis}
\end{split}\end{equation}
In this basis the concurrence can be written as
\begin{equation}C=\mathrm{max}\left\{0,\lambda_1-\lambda_2-\lambda_3-\lambda_4\right\},
\label{concurrence} \end{equation}
where $\lambda_i$ are the decreasingly ordered square roots of the eigenvalues of $\rho_{\mathrm{magic}}\times(\rho_{\mathrm{magic}})^{\ast}$, with $\rho_{\mathrm{magic}}$ being the reduced density matrix in the magic basis \cite{Wootters}. After some algebra (see App. \ref{sec:concurrence} for details) we end up with the following formula for the concurrence:
\begin{equation}
C=4\max\left\{0,H+2I\right\}.
\end{equation}
This implies that the concurrence can actually be expressed in terms of a \emph{single} parameter
\begin{equation}C=\max\{0,\mathcal{E}\},\label{concisnozero2}
\end{equation}
where
\begin{equation}\mathcal{E}=\frac{4}{3}\left(\langle S_z^1S_z^2\rangle-\langle S_x^1S_x^2\rangle-3\langle S_y^1S_y^2\rangle\right).\label{eq:singleparameter}\end{equation}

\begin{figure}
\epsfig{file=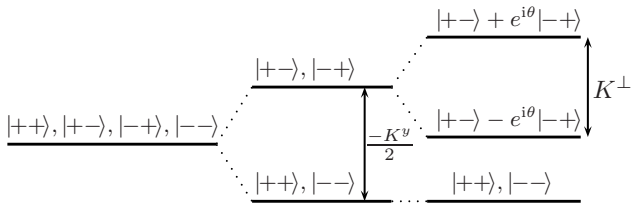, scale=.95}
\centering
\caption{Level splitting due to the RKKY interaction.}\label{fig:rkkylevels}
\end{figure}

To calculate the concurrence of any state of the two-spin subsystem (arising from a total Hamiltonian conserving $U(1)$ spin symmetry) we can use the formula (\ref{concisnozero2}) derived above.
Deep in the RKKY regime where the competition from the Kondo interaction can be neglected (the case discussed in Sec.\! \ref{section:RKKY}), the spins are described by a Hamiltonian of the form
\begin{equation}{H}=K^y S_1^yS_2^y+\frac{1}{2}e^{\i \theta}K^{\perp}(S_1^z + iS_1^x)(S_2^z - iS_2^x) +h.c. \end{equation}
which has three possible ground states: a rotated singlet state
\begin{equation}\rho_\text{singlet}(\theta)\equiv\frac{1}{2}\left(|\hspace{-1mm}+\hspace{-.75mm}-\hspace{-.25mm}\rangle-e^{i\theta}|\hspace{-1mm}-\hspace{-.75mm}+\hspace{-.25mm}\rangle\right)\left(\langle\hspace{-.25mm}+\hspace{-.75mm}-\hspace{-1mm}|-e^{-i\theta}\langle\hspace{-.25mm}-\hspace{-.75mm}+\hspace{-1mm}|\right),\label{eqn:rhosinglet}\end{equation} an Ising ground state 
\begin{equation}
\rho_\text{Ising} =\frac{1}{2}|\hspace{-1mm}+\hspace{-.75mm}+\hspace{-.25mm}\rangle\langle\hspace{-.25mm}+\hspace{-.75mm}+\hspace{-1mm}|+\frac{1}{2}|\hspace{-1mm}-\hspace{-.75mm}-\hspace{-.25mm}\rangle\langle\hspace{-.25mm}-\hspace{-.75mm}-\hspace{-1mm}|,\label{eqn:rhodoublet}
\end{equation} 
and a mixture of rotated triplet states
\begin{equation}
\rho_\text{triplet}(\theta)=\frac{1}{3}\rho_\text{singlet}(\theta)+\frac{1}{3}|\hspace{-1mm}+\hspace{-.75mm}+\hspace{-.25mm}\rangle\langle\hspace{-.25mm}+\hspace{-.75mm}+\hspace{-1mm}|+\frac{1}{3}|\hspace{-1mm}-\hspace{-.75mm}-\hspace{-.25mm}\rangle\langle\hspace{-.25mm}-\hspace{-.75mm}-\hspace{-1mm}|,\label{eqn:rhotriplet}
\end{equation}
corresponding to the cases $K^\perp>-K^y$, $K^\perp<-K^y$ (or $|K^y|>K^\perp=0$) and $K^\perp=-K^y$, respectively. This can be read off immediately from the level scheme in FIG. \ref{fig:rkkylevels}. Note that the states in Eqs. (\ref{eqn:rhosinglet}) and (\ref{eqn:rhotriplet}) are just the conventional singlet ($\theta=0$) and triplet states ($\theta=\pi $) in a basis where one of the spins has been rotated with respect to the other (cf. Eq. (\ref{eqn:rotatedrkky})). 
The parameter $\mathcal{E}$ that determines the concurrence in those three cases is (cf. Eq. (\ref{concisnozero2})
\begin{equation}\begin{array}{llll}
&\mathcal{E}_\text{singlet}&=+1\\
&\mathcal{E}_\text{Ising}&=-1/3\\
&\mathcal{E}_\text{triplet}&=-1,
\end{array}
\end{equation}
hence the concurrence is unity for the singlet while it always vanishes in the latter two cases (see Appendix \ref{sec:concurrence} for details). 

To understand what effects a spin-orbit interaction my have on the entanglement in the RKKY regime, consider the situation described by the $F_0$-amplitude in Eq. (\ref{eqn:ffunctions}), i.e. an RKKY interaction given by
\begin{equation}H=-\frac{J^2}{2\pi^2R^2}\frac{m^2}{\hbar^2}\sin 2R\sqrt{k_F^2+\frac{m^2\alpha^2}{\hbar^4}} \vect{S}_1\cdot\vect{S}_2.\end{equation}
It is clear from this that a change in $\alpha$ may change the overall sign of the interaction and thus drive a system with a maximally entangled singlet ground state to one with a non-entangled triplet ground state, or vice versa. As an illustration, one may consider a double-quantum dot structure patterned in an InAs heterostructure, which, due to its large electron mean-free path, is favored in most spintronics applications \cite{Zutic}. Using data for a heterostructure grown by molecular beam epitaxy \cite{Grundler}, with a gate controlled Rashba parameter tunable in the range $1 - 5 \times 10^{-11}$ eV m, and with an effective mass $m \approx 0.4 m_e$, one finds that the inverse Rashba  spin-orbit length $m \alpha /\hbar^2$ varies in the interval  $0.01 - 0.1$ \AA $^{-1}$ (to be compared to the Fermi wave number $k_F \sim 0.02$  \AA $^{-1}$, and the distance $R$ between the dots, say, of the order of magnitude $R \sim 10^2$ \AA  ). In the more general case $-$ when the $F_1$- and $F_2$-amplitudes in Eq. (\ref{eqn:ffunctions}) come into play $-$ it must be expected that a transition to a system with a non-entangled Ising ground state can also occur. We conclude that spin-orbit interactions may drastically reduce, or, enhance, the entanglement between two spin qubits, with the precise effect depending on materials and design parameters, as well as on the strength of applied
electrical fields. It is here important to emphasize that estimating the value of the RKKY coupling in a real device is no easy task. In addition to the uncertainty in determining the distance between the two spin qubits, there are difficulties in judging the influence from the mean-free path, dephasing length, etc. An exquisitely precise control of the experimental set-up is a {\em sine qua non} for a reliable estimate of the RKKY coupling  on which entanglement depends so crucially.  

\section{Entanglement at criticality}\label{entanglementdiscussion}

Using the formalism developed in the previous section (see Appendix \ref{sec:concurrence} for details), and exploiting the results of Sec. \ref{sec:sofix} we find (cf. Eq. (\ref{eq:singleparameter}))
\begin{equation}\mathcal{E}_{\text{critical}}=0,\end{equation}
i.e. the concurrence vanishes at the critical point $K^\perp=K^y\approx 2.2 T_K$ as in the $SU(2)$ symmetric case with no spin-orbit interaction \cite{cho:012109}. It should be noted that in the cases of a vanishing concurrence in the noncritical RKKY regime, i.e. for the Ising doublet and the triplet states (discussed in Sec. IV), the value of $\mathcal{E}$ was always a finite and negative number, and the concurrence vanished by virtue of taking the maximum of zero and this number. In the case of the critical ground state, $\mathcal{E}$ is by itself zero, which means that a small deviation from the critical point could lead to a non-zero value for the concurrence. In other words, the point where the concurrence first vanishes marks the critical point. Here a stronger antiferromagnetic coupling drives the system to the free fixed point where the spins form a fully entangled singlet, while a weaker antiferromagnetic or even ferromagnetic coupling drives the system to the fully screened two-channel Kondo fixed point where there is no entanglement between the spins.

It is important to keep in mind that the BCFT result in Ref. \onlinecite{PhysRevB.52.9528} about the irrelevance of anisotropies is only valid for small deviations around the isotropic case. In the device which we suggest (cf. FIG. 3), we have shown that the anisotropy in the $xz$-plane can be arbitrarily large, i.e. the phase of $K$ does not change the picture. For large anisotropies along the $y$-axis this will be quite different. We are not able to describe the exact behavior in the entire parameter plane spanned by $K^\perp$ and $K^y$, but for certain regions there are strong arguments that the system should behave in a certain way. Let us first consider the case $|K^y|\ll K^\perp$. For very large values of $K^\perp$ the impurity spins will form a singlet and the conduction electrons are decoupled, just as for $K^y\approx K^\perp \gg T_K$. For $K\rightarrow 0$ (and $|K^y|< K^\perp$) the system will flow to the two-channel Kondo fixed point, again just like in the isotropic case. In fact, the same line of arguments used in Ref. \onlinecite{gan} can be applied. The impurity state at the critical point should be the same, in particular also having vanishing concurrence. We expect that this fixed point can be connected to the isotropic fixed point by tuning both $K^y$ and $K^\perp$ appropriately, and extended for larger values of $K^\perp >-K^y>0$.
For $|K^y|\gg K^\perp, |K_y|\gg T_K$ we are in the RKKY regime where the impurity spins are completely decoupled from the conduction electrons. We thus know that $K^y\ll 0$ gives vanishing concurrence, while for $K^y\gg 0$ a nonzero value of $K^\perp$ raises the concurrence from $0$ to $1$. There is, however, an intermediate region $K^y\sim T_K, K^\perp=0$ where a different quantum phase transition is expected to take place \cite{isingcoupled}. An analysis of the crossover to this quantum critical point, or of the entanglement properties in its vicinity, is beyond the scope of this paper. 

\section{Summary}

We have carried out a systematic study of how spin-orbit interactions $-$ of Rashba as well as of Dresselhaus type $-$ influence the Kondo and RKKY interactions in a nanoscale device described by a two-dimensional two-impurity Kondo model. By absorbing the Rashba and Dresselhaus interactions as anisotropies of an effective RKKY coupling we can monitor their effects for varying coupling strengths. This provides us with a controlled inroute to study the quantum critical behavior of the model in presence of spin-orbit interactions, exploiting known results from boundary conformal field theory \cite{PhysRevB.52.9528} and an effective bosonized Hamiltonian approach \cite{gan}. Most strikingly, and as illustrated in Fig. 4, for a particular combination of symmetry breaking terms there is strong evidence for a {\em line of critical points} exhibiting the same universal behavior as that of the known isotropic model (up to RG irrelevant scaling corrections). It remains a challenge to rigorously establish the existence of this critical line.  We have also studied how spin-orbit interactions influence the entanglement between two RKKY-coupled spinful quantum dots ({\em alias} ''qubits'' in the language of quantum computing). Using data for a nanoscale device patterned in a gated InAs heterostructure we find that a gate-controlled Rashba spin-orbit interaction may drive a maximally entangled state to one with vanishing entanglement, or vice versa (as measured by the two-qubit concurrence). This has important implications for proposals using RKKY interactions for nonlocal control of qubit entanglement in semiconductor heterostructures. 

{\bf Acknowledgements.} We wish to thank H. R. Krishna-murthy and Y. Meir for discussions and helpful correspondence. This work was supported by the Swedish Research Council under grant No. VR-2005-3942.
\appendix
\section{RKKY interaction for $\hat{R}=\hat{x}$} \label{app:spintraces}
To derive the form of the RKKY interaction, it is necessary to perform traces over various combinations of Pauli matrices
$\tau^x, \tau^y$, and $\tau^z$. Specifically, we need the relations\small
\begin{eqnarray}
\text{Tr}\left[\left(\vect{\tau}\cdot \vect{A}\right)\left(\vect{\tau}\cdot \vect{B}\right)\right]&=&2 \vect{A}\cdot\vect{B}, \nonumber \\
\text{Tr}\left[\left(\vect{\tau}\cdot \vect{A}\right)\left(\vect{\tau}\cdot \vect{B}\right) \tau^i\right]&=&2\i (\vect{A}\times\vect{B})_i, \nonumber \\
\text{Tr}\left[\left(\vect{\tau}\cdot \vect{A}\right)\tau^j\left(\vect{\tau}\cdot \vect{B}\right) \tau^i\right]&=&2A_iB_j+2A_jB_i-2 \delta_{ij}\vect{A}\cdot\vect{B}, \nonumber
\end{eqnarray}
\normalsize
which can be easily verified using \small
\begin{equation}\left(\vect{\tau}\cdot \vect{A}\right)\left(\vect{\tau}\cdot \vect{B}\right)=\left(\vect{A}\cdot \vect{B}\right)+\i\left(\vect{A}\times \vect{B}\right)\cdot \vect{\tau}.\end{equation}\normalsize
For the case $\hat{R}=\hat{x}$, discussed in the text after Eq. (\ref{eq:G1}), the single-electron Green's function is
\begin{equation}G=G_0+G_1\alpha \,\tau^y-G_1\beta\, \tau^x \label{G} \end{equation}
The RKKY interaction is proportional to
\begin{equation}\label{Tr} \text{Tr}\left[(\vect{S}_1\cdot\vect\tau)G(\vect{R})(\vect{S}_2\cdot\vect\tau)G(-\vect{R})\right].\end{equation} 
Inserting $G$ from Eq. (\ref{G}) into (\ref{Tr}) and using the trace formulas given above, we obtain:
\begin{equation}
 \begin{array}{rlll}
 \text{Tr}\big[(\vect{S}_1&\!\!\!\!\cdot\vect\tau)G(\vect{R})(\vect{S}_2\cdot\vect\tau)G(-\vect{R})\big]=\\[2mm]
 & 2 G_0^2 \vect{S}_1\cdot\vect{S}_2  \\[1mm]
 -&  4G_0G_1\i\alpha\left(\vect{S}_1\times\vect{S}_2\right)^y  \\[1mm]
+&4G_0G_1\i\beta\left(\vect{S}_1\times\vect{S}_2\right)^x \\[1mm]
+&  2G_1^2\alpha^2\left(\vect{S}_1\cdot\vect{S}_2-2S_1^zS_2^y\right)  \\[1mm]
 +&  2G_1^2\beta^2\left(\vect{S}_1\cdot\vect{S}_2-2S_1^xS_2^x\right) \\[1mm]
 +&  4G_1^2\alpha\beta \left(S_1^xS_2^y+S_1^yS_2^x\right)  .
\end{array}
\end{equation}

\section{Concurrence for various states} \label{sec:concurrence}

The concurrence for two qubits (two-spin subsystem) can be written as
\begin{equation}C=\mathrm{max}\left\{0,\lambda_1-\lambda_2-\lambda_3-\lambda_4\right\},
\label{concurrence2} \end{equation}

where $\lambda_i$ are the decreasingly ordered square roots of the eigenvalues of $\rho_{\mathrm{magic}}\times\rho_{\mathrm{magic}}^{\ast}$, with $\rho_{\mathrm{magic}}$ being the reduced density matrix in the magic basis  \cite{Wootters}. In terms of $H, I, D$, and $X$ defined in Eq. (\ref{eqn:parameters}), it is given by
\begin{equation}\rho_{\mathrm{magic}}=\begin{pmatrix}-H'-2I&-\i X^+&0&0\\\i X^+&-H'-2I&0&0\\0&0&4I-H'&D+\i X^-\\0&0&D-\i X^-&\frac{1}{4}+3H\end{pmatrix} \label{eqn:rhomagical}\end{equation}
where $H' \equiv H-1/4$.
Since $\rho_{\mathrm{magic}}\times\rho_{\mathrm{magic}}^{\ast}$  is block diagonal it is straightforward to calculate the eigenvalues and one finds

\begin{equation}\begin{split}
&\ \lambda^2_{1,2}=\left(\frac{1}{4}-H-2I\right)^2-(X^+)^2\\
&\lambda^2_{\pm}= D^2-(X^-)^2+(2I-2H)^2+\left(2I+H+\frac{1}{4}\right)^2\\&\ \ \pm 2\sqrt{((2I-2H)^2+D^2) \left(\left(2I+H+\frac{1}{4}\right)^2-(X^-)^2\right)}.
\end{split} \label{eigenvalues} \end{equation}
\normalsize
Note that the parameters are not fully independent; all eigenvalues are positive numbers.

To calculate the concurrence by the formula in (\ref{concurrence2}), it is necessary to order the eigenvalues by size.
Since $\lambda_-$ can never be the largest eigenvalue, there are only two cases, $\lambda_+ >\lambda_1$ and $\lambda_1>\lambda_+$, to analyze.
First consider $\lambda_+>\lambda_1$.
In this case Eq. (\ref{concurrence}) takes the form
\begin{equation}
C=\max\{0,\lambda_+-\lambda_--2\lambda_1\},\label{concisnozero3}
\end{equation}
which, in particular, implies that the concurrence vanishes for
$\lambda_+-\lambda_-\leq 2\lambda_1$.
As for the second possibility, $\lambda_1>\lambda_+$, inspection of  Eq. (\ref{concurrence2}) reveals that the concurrence vanishes identically for this case since $\lambda_1-\lambda_2=0$ and $\lambda_{\pm}\geq 0$. However, $\lambda_+-\lambda_-\leq \lambda_+\leq \lambda_1\leq 2\lambda_1$,
thus the concurrence also vanishes by Eq. (\ref{concisnozero3}). It follows that the second case is already contained in that formula and it is not necessary to treat the two cases separately.

In the following we shall specialize to the case with $X^{\pm}=0$ (cf. Section V) in which case the $\lambda_i$ simplify to
\begin{equation}
\begin{split}
\lambda_{1}& = \frac{1}{4}-H-2I\\
\lambda_{\pm}&=\sqrt{D^2+\left(2I-2H\right)^2}\pm\left|2I+H+\frac{1}{4}\right|,
\end{split}\label{eqn:lambdas}
\end{equation}
and we immediately end up with Eq. (\ref{concisnozero2}), i.e.
\begin{equation}
\begin{split}
C&=\max\left\{0,\left|4I+2H+\frac{1}{2}\right|-\frac{1}{2}+2H+4I\right\}\\\label{app:concurrenceformula}
&=4\max\left\{0,H+2I\right\}\\
&=\max\left\{0,\mathcal{E}\right\}.
\end{split}
\end{equation}

To calculate the concurrence for a given reduced density matrix in the magic basis, as parameterized in Eq. (\ref{eqn:rhomagical}), we need to calculate the expectation value $\mathcal{E}$ as defined in Eq. \ref{eq:singleparameter}. The corresponding operator takes the following form in the magic basis:
\begin{equation}\begin{split}
\mathcal{O}_\mathcal{E}&=\frac{4}{3}\left(\langle S^z_1S^z_2\rangle-\langle S^x_1S^x_2\rangle-3\langle S^y_1S^y_2\rangle\right)\\
&=\frac{1}{3}\mathrm{diag}(-1,-5,3,3)\label{eq:tangleoperator}
\end{split}\end{equation}

Deep in the RKKY regime there are only three possible states for the two coupled qubits:, a singlet, an Ising doublet, and a triplet. Starting with the singlet state, (Eq. \ref{eqn:rhosinglet}), its density matrix in the magic basis is given by
\begin{equation}\rho_{\text{singlet}}(\theta)=\mathrm{diag}\left(\begin{pmatrix}0&0\\0&0\end{pmatrix},\begin{pmatrix}\hfill \sin^2\frac{\theta}{2}&\frac{1}{2}\sin\theta\\\frac{1}{2}\sin\theta&\hfill \cos^2\frac{\theta}{2}\end{pmatrix}\right),\end{equation}
and $\mathcal{E}$ is determined as
\begin{equation} \begin{split}\label{HID}
\mathcal{E}_\text{singlet}(\theta)&=\mathrm{Tr}[\mathcal{O}_\mathcal{E}\rho_\text{singlet}(\theta)]\\&=\frac{1}{3}\mathrm{Tr}\left[\begin{pmatrix}\hfill 3\sin^2\frac{\theta}{2}&\frac{3}{2}\sin\theta\\ \frac{3}{2}\sin\theta&\hfill 3\cos^2\frac{\theta}{2}\end{pmatrix}\right] = 1.\end{split}
\end{equation}
This means that the concurrence for the singlet state is unity, independent of $\theta$. In other words, the well-known fact that a singlet state is maximally entangled is immediately reproduced in our formalism. The Ising doublet, (Eq. \ref{eqn:rhodoublet}), has the density matrix
\begin{equation}\rho_{\text{Ising}}=\text{diag}(1/2,1/2,0,0).\end{equation}
By inspection $\mathcal{E}=\frac{1}{3}\times-\frac{6}{2}=-1$, thus the concurrence is zero for this case.
As for the triplet state, Eq. (\ref{eqn:rhotriplet}), the parameters for $\rho_\text{triplet}(\theta)$ can easily be obtained from the ones for $\rho_\text{singlet}(\theta)$ and $\rho_\text{Ising}$, since these density matrices as well as the spin-operators are block diagonal. One finds
\begin{eqnarray}
\mathcal{E}_\text{triplet}(\theta)&=&\frac{1}{3} \mathcal{E}_\text{singlet}(\theta)+\frac{2}{3} \mathcal{E}_\text{Ising}=-\frac{1}{3}, \nonumber 
\end{eqnarray}
and the concurrence is again seen to vanish identically.

Turning to the critical ground state of the TIKM, its reduced density matrix for the two qubits is given by Eq. (\ref{rhokondo}), with a non-zero value of $\theta$ when spin-orbit interactions are present, 
\begin{equation}
\rho_\text{critical}(\theta) =\frac{1}{4}\begin{pmatrix}1 & 0 & 0 & 0 \\ 0 & 1 & 0 & 0 \\ 0 & 0 &2\sin^2\frac{\theta}{2}&\sin\theta \\0 & 0 &\sin\theta&\hfill2 \cos^2\frac{\theta}{2} \label{rhokondo} \end{pmatrix}.
\end{equation}
From this we find that $\mathcal{E}_\text{critical}=0$, i.e. the concurrence vanishes at the critical point.\\

\section{2D Kondo model in the presence of Rashba or Dresselhaus interaction}\label{app:sokondo}
The Hamiltonian for Rashba and Dresselhaus spin-orbit interactions in two dimensions, which we take as the $x$-$y$ plane, is (see Eqs. (\ref{Rashba}) and (\ref{Dresselhaus2D})) 
\begin{equation}H_{\text{s-o}}=\left(\beta k_x-\alpha k_y\right)\tau^x-\left(\beta k_y-\alpha k_x\right)\tau^y.\end{equation}

We can define raising and lowering operators for spin and angular momentum as
\begin{equation}\begin{split}\tau^\pm&=\frac{1}{2}(\tau^x-\i\tau^y)\\
L^{\pm}&=k_x\pm \i k_y,
\end{split}\end{equation}
to rewrite the Hamiltonian as
\begin{equation}H_{\text{s-o}}=\alpha \tau^-L^++\beta \tau^+L^++h.c.\end{equation}
This expression suggests to write the free fermion fields $\psi_{\vect{k}}$ in terms of an angular momentum quantum number $m$ and the magnitude $k$ of the momentum as \cite{arXiv:0708.2435}
\begin{equation}
\psi_{km}=\sqrt{\frac{k}{2\pi}}\int_0^{2\pi}d\theta e^{-\i m \theta} \psi_{\vect{k}}.
\end{equation}
The second-quantized spin-orbit Hamiltonian can then be expressed as
\begin{multline}H_{\text{s-o}} =\int dk  \sum_{m,\sigma,\sigma'}k^2(\alpha \psi^{\dagger\sigma}_{k,m+1}(\tau^+)_\sigma^{\sigma'}\psi_{k,m,\sigma'} \\ -\i\beta \psi^{\dagger\sigma}_{k,m-1}(\tau^-)_\sigma^{\sigma'}\psi_{k,m,\sigma'})+h.c\label{eqn:app:h2nd:final}.\end{multline} \\
It follows that the 2D Hamiltonian for the Kondo model with added Rashba and Dresselhaus spin-orbit interactions takes the form 
\begin{equation}
H=H_\text{kin}+H_\text{Kondo}+H_{\text{s-o}}\label{eqn:app:kondo}
\end{equation}
where \small
\begin{equation}\begin{split}
H_\text{kin}&=\int dk  \mathop{\sum_m}_{\sigma=\uparrow,\downarrow} k\epsilon_k\psi^{\dagger\sigma}_{k,m}\psi_{k,m,\sigma}\\
H_\text{Kondo}&=J \int dk \int dk' \sqrt{kk'} \mathop{\sum_{\sigma,\sigma'=\uparrow,\downarrow}}\psi^{\dagger \sigma}_{k,0}\vect{\tau}_{\sigma}^{\sigma'}\psi_{k',0,\sigma'}\cdot \vect{S}.
\end{split}\end{equation}
\normalsize
and where $H_{\text{s-o}}$ is given in Eq. (\ref{eqn:app:h2nd:final}).

As manifest in Eq. (\ref{eqn:app:h2nd:final}), when both kinds of spin-orbit interactions are present, all $m$-states couple to the impurity.
On the other hand, if only one type of spin-orbit interaction is present, there are just two states coupling to the impurity.
To diagonalize the spin-orbit Hamiltonian in this case, we define new fields as
\begin{equation}\begin{split}\psi_{k,m,\uparrow,\pm}&=\frac{1}{\sqrt{2}}\left(\psi_{k,\uparrow, m}\pm \mbox{e}^{i\pi(n-1)/2} \psi_{k,\downarrow, m+n}\right)\\
\psi_{k,m,\downarrow,\pm}&=\frac{1}{\sqrt{2}}\left(\psi_{k,\downarrow, m}\pm \mbox{e}^{i\pi(n-1)/2}  \psi_{k,\uparrow ,m-n}\right),\end{split}\label{newfields}\end{equation}
where $n=1$ in the Rashba case $(\beta=0)$ and $n=-1$ in the Dresselhaus case $(\alpha=0)$, and where the $\uparrow, \downarrow$ labels on the new fields refer to the ''pseudospin'' $\bar{\sigma}$ introduced in FIG. (4).

In terms of these fields, the three terms of the Hamiltonian in (\ref{eqn:app:kondo}) are given by 
\begin{equation}
\begin{split}
H_\text{kin}&=\int dk \sum_m \mathop{\sum_{\bar{\sigma}=\uparrow\downarrow}}_{f=\pm} \epsilon_{k}\psi^{\dagger\bar{\sigma}}_{k,m,f}\psi_{k,m,\bar{\sigma},f}, \\
H_\text{Kondo}&=\frac{J}{2}\! \int \!dk\! \int \!dk' \sqrt{kk'} \mathop{\sum_{\bar{\sigma},\bar{\sigma}'=\uparrow,\downarrow}}_{f,f'=\pm}\hspace{-1mm} \psi^{\dagger \bar{\sigma}}_{k,0,f}\vect{\tau}^{\bar{\sigma}'}_{\bar{\sigma}}\psi_{k',0,\bar{\sigma}',f'}\cdot \vect{S}, \\
H_\text{s-o}&=\mu\int dk k\sum_{m} \mathop{\sum_{\bar{\sigma}=\uparrow,\downarrow}}_{f=\pm} f\,\psi^{\dagger\bar{\sigma}}_{k,m,f}\psi_{k,m,\bar{\sigma},f}, 
\end{split}
\end{equation} 
where $\mu\!=\!\alpha$ if the spin-orbit interaction is of the Rashba type and $\mu\!=\!\beta$ if it is of the Dresselhaus type.
Since all terms are diagonal in $m$ and only the $m\!=\!0$ fields take part in the Kondo interaction, we drop all other fields and suppress the $m$ quantum number which is always zero from here on. Defining $\epsilon_k^\pm=\epsilon_k\pm \mu k$, we combine $H_\text{kin}$ and $H_{\text{s-o}}$ to
\begin{equation}\tilde{H}_\text{kin}=\int dk \mathop{\sum_{\bar{\sigma}=\uparrow,\downarrow}}_{f=\pm}\hspace{-1mm} \epsilon^f_k \psi^{\dagger\bar{\sigma}}_{k,f}\psi_{k,\bar{\sigma},f}. \end{equation}
As a last step we define a set of fields as
\begin{equation}\psi_{E,f,\bar{\bar{\sigma}}}=\frac{\int dk \sqrt{\frac{k}{2}}\delta(E-\epsilon^f_k)\psi_{k,\bar{\sigma},f}}{\int dk' \delta(E-\epsilon^f_{k'})},\end{equation}
to obtain the Hamiltonian
\begin{multline}H= \mathop{\sum_{\bar{\sigma},\bar{\sigma}'=\uparrow,\downarrow}}_{f,f'=\pm} \int dE E \psi^{\dagger\bar{\sigma}}_{E,f}\psi_{E,f,\bar{\sigma}} \\
+\frac{1}{2}\int dE\int dE'  \mathop{\sum_{\bar{\sigma},\bar{\sigma}'=\uparrow,\downarrow}}_{f,f'=\pm}\hspace{-1mm} J^{f,f'}(E,E')\psi^{\dagger\bar{\sigma}}_{E,f}\vect{\tau}_{\bar{\sigma}}^{\bar{\sigma}'}\psi_{E',f',\bar{\sigma}'}
 \cdot\vect{S},\end{multline}
where
\begin{equation}J^{f,f'}(E,E')=J \int dk \delta(E-\epsilon^f_{k})\int dk' \delta(E'-\epsilon^{f'}_{k'}).
\end{equation} 


\begin{thebibliography}{99}
\bibitem{FootnoteOnOrder} 
The physics does not, of course, depend on the order in which the system is set up. A similar argument holds for starting with the RKKY interaction and then adding the Kondo screening channels.
\bibitem{footnote3}
The marginal operator ${\cal O}$ that becomes allowed in the presence of a Dzyaloshinski-Moriya
interaction can be written as ${\cal O} = (h_1^\dagger)^A(h_2)_A\phi^y$,
where $h_i$ are isospin-$SU(2)_1$ WZW-fields with spin $1/2$ and $\vect{\phi}$ is a spin-$SU(2)_2$ WZW-field with spin $1$ \cite{PhysRevB.52.9528}.
\bibitem{NotCritical}
Conversely, a system that is not critical at first may become critical upon introducing spin-orbit interactions.
\bibitem{ParticleHole}
 It should be kept in mind that in the presence of charge transfer, the fixed point is stable only in the case  
 of particle-hole symmetry \cite{PhysRevB.52.9528}.
\end{thebibliography}
\end{document}